\documentclass[%
 aip,
 jcp,%
 amsmath,amssymb,
preprint,%
reprint,%
]{revtex4-1}

\usepackage{graphicx}% Include figure files
\usepackage{dcolumn}% Align table columns on decimal point
\usepackage{bm}% bold math
\begin{document}

%\preprint{AIP/123-QED}

%\title[]{Charge transport through molecular junctions: from Landauer--B\"{u}ttiker to Marcus theory and beyond}% Force line breaks with \\

\title[]{Beyond Marcus theory and the Landauer-B\"{u}ttiker approach in molecular junctions: A unified framework }

\author{Jakub K. Sowa}
 \email{jakub.sowa@materials.ox.ac.uk}
\affiliation{ 
Department of Materials, University of Oxford, OX1 3PH Parks Road, Oxford, UK}%
\author{Jan A. Mol}
\affiliation{ 
Department of Materials, University of Oxford, OX1 3PH Parks Road, Oxford, UK}%
\author{G. Andrew D. Briggs}
\affiliation{ 
Department of Materials, University of Oxford, OX1 3PH Parks Road, Oxford, UK}%
\author{Erik M. Gauger}
\affiliation{ 
SUPA, Institute of Photonics and Quantum Sciences, Heriot-Watt University, Edinburgh EH14 4AS, UK}%

\date{\today}% It is always \today, today,

\begin{abstract} %should be approx 250 words
Charge transport through molecular junctions is often described either as a purely coherent or a purely classical phenomenon, and described using the Landauer--B\"{u}ttiker formalism or Marcus theory, respectively. 
Using a generalised quantum master equation, we here derive an expression for current through a molecular junction modelled as a single electronic level coupled to a collection of thermalised vibrational modes. We demonstrate that the aforementioned theoretical approaches can be viewed as two limiting cases of this more general expression, and present a series of approximations of this result valid at higher temperatures. We find that Marcus theory is often insufficient in describing the molecular charge transport characteristics and gives rise to a number of artefacts, especially at lower temperatures. Alternative expressions, retaining its mathematical simplicity but rectifying those shortcomings, are suggested.
In particular, we show how lifetime broadening can be consistently incorporated into Marcus theory, and we derive a low-temperature correction to the semi-classical Marcus hopping rates.
Our results are applied to examples building on phenomenological as well as microscopically-motivated electron-vibrational coupling.
We expect them to be particularly useful in experimental studies of charge transport through single-molecule junctions as well as self-assembled monolayers.
\end{abstract}

\maketitle

\section{\label{Intro}Introduction}
In the last twenty years single-molecule electronics has transformed from an exotic to a well-established, fast-developing field.\cite{nitzan2003electron,aradhya2013single} This transition has been predominantly driven by enormous technological progress in the fabrication of single-molecule junctions (SMJs).
These devices comprise an individual molecule spanning a gap between two metallic electrodes. Such a setup allows for the passing and measuring of the electric current flowing through the studied structure.
There currently exists a number of techniques which can be used to fabricate SMJs. Historically, methods utilising Scanning Tunnelling Microscopy have perhaps  been the most important.\cite{cui2001reproducible,venkataraman2006dependence,stipe1998single,sedghi2011long} Over the years, various other techniques have been developed, based on: break-junctions,\cite{reed1997conductance,bohler2004mechanically,perrin2014large} electro-migrated gold electrodes,\cite{park2000nanomechanical,van2006molecular} electroburnt\cite{prins2011room,mol2015graphene,gehring2017distinguishing,burzuri2016sequential} and etched graphene nano-junctions.\cite{jia2016covalently,jia2013conductance,xu2017single}
Some of these device geometries feature a so-called gate electrode, which allows for  electrostatic control of the molecule. This enables operation in the resonant transport regime, where the molecular energy levels lie within the bias window, and the non-resonant regime, where they are outside it.

Electron-vibrational (electron-phonon) interactions can play a significant role in charge transport through molecular junctions.\cite{galperin2007molecular} In the off-resonant regime these effects are typically relatively modest but they can have an enormous influence on the resonant transport characteristics.
The theory of vibrational effects in resonant transport is by now quite well developed. Intermediate and strong coupling to individual molecular vibrational modes typically gives rise to steps in the $IV$ characteristics (or peaks in the differential conductance).\cite{riss2014imaging,braig2003vibrational,flensberg2003tunneling,park2000nanomechanical} It has further been shown that these interactions can result in a breadth of other phenomena including: negative differential conductance, rectification and current blockade (known as the Franck--Condon blockade).\cite{koch2005franck,koch2006theory,hartle2011vibrational,hartle2011resonant,galperin2005hysteresis,zazunov2006phonon}
Interactions with a collection of thermalised modes (weakly coupled molecular modes or modes in the solvent or the substrate) do not induce similarly clear signatures in the current-voltage characteristics. Their influence is usually studied by considering the temperature dependence of the transport behaviour.\cite{kilgour2015charge,choi2010transition,taherinia2016charge,segal2002conduction,nitzan2006chemical}

In spite of the advancements in the field, experimental measurements of charge transport through molecular junctions remain challenging. The main issue continues to be the reproducibility of results between different junctions comprising the same molecular structures.
This problem is inherent to single-molecule measurements and stems mainly from  differences in the microscopic structure of the leads, variability in the molecule-lead contacts, and geometric distortions of the deposited molecular structures.
As the result, there currently exists a dissonance between the theoretical modelling of (inelastic) resonant transport and the majority of experimental studies on the topic. 
The analysis in the latter is often limited to qualitatively identifying the vibrational features,\cite{burzuri2016sequential,repp2010coherent,osorio2010conductance,pasupathy2005vibration} as reproducing the full $IV$ characteristics has often proven to be problematic (with a few notable exceptions\cite{secker2011resonant,riss2014imaging}).
This is clearly unsatisfactory, and therefore there exists a need for a simple theoretical framework which captures the relevant transport phenomena (beyond single-vibrational-mode models) at the minimal required level of complexity.

The objective of this work is to arrive at an expression for the steady-state electric current through a weakly-coupled molecular junction in the resonant transport regime which will capture the effects of vibrational coupling and lifetime broadening, and also account for the charge state of the molecular system. We shall achieve this using a relatively simple generalised quantum master equation.\cite{esposito2009transport}
Although there currently exist a number of sophisticated theoretical approaches that have achieved the goal set out above (most notably methods using Hierarchical Equations of Motion),\cite{wang2011numerically, white2012inelastic,schinabeck2016hierarchical,dou2018broadened} their complexity typically restricts their use in explaining the experimental measurements.  
The second goal of this work is to derive a number of approximate expressions valid at higher temperatures which can be very easily computed or perhaps even fitted to empirical data. 
We will focus on vibrational effects in short molecular junctions (modelled as a single site) although the crucial role of electron-phonon coupling in charge transport through longer molecular wires, DNA, and similar structures has also been demonstrated. \cite{zimbovskaya2007dissipative,gutierrez2005dissipative,kim2017controlling,kilgour2015charge,segal2002conduction,sowa2017vibrational,zimbovskaya2018thermally}
\begin{figure}
    \centering
    \includegraphics{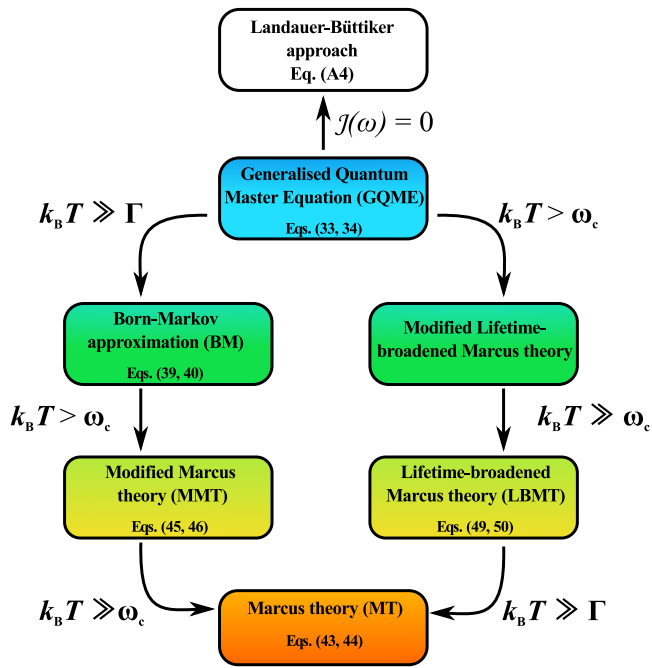}
    \caption{Schematic illustration of the main conceptual result of this work. $T$ denotes the temperature of the system, $k_{\mathrm{B}}$ is the Boltzmann constant, $\Gamma$ is the lifetime broadening, and $\omega_c$ is the cut-off frequency of the phonon bath.}
    \label{scheme1}
\end{figure}

This work is organised as follows.
In Section \ref{model}, we outline the theoretical model used in this study. Subsequently, in Section \ref{GQ}, we derive and discuss a compact expression for the electric current flowing through the junction. As mentioned above, it will be obtained using the generalised quantum master equation (GQME) in the polaron-transformed frame. Section \ref{highT} discusses approximate expressions of this result valid in various parameter regimes. As we shall demonstrate, a number of simplifications of the expression from Section \ref{GQ} can be obtained for increasing temperature resulting in a tiered theoretical approach, schematically pictured in Fig.~\ref{scheme1}.
These results are further discussed and summarised in Section \ref{end}.

\section{Model \label{model}}
\begin{figure}
    \centering
    \includegraphics{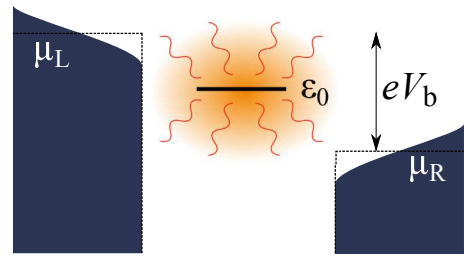}
    \caption{A graphical illustration of the system studied in this work. We model the molecular junction as a single electronic level coupled to a collection of thermalised vibrational modes, as well as the left (L) and right (R) electrode. $V_\mathrm{b}$ denotes the bias potential, $\mu_\mathrm{L}$ and $\mu_\mathrm{R}$ are the chemical potentials of the left and right electrode, respectively.}
    \label{scheme2}
\end{figure}

Our theoretical model is schematically pictured in Fig.~\ref{scheme2}, and described by the following Hamiltonian (we set $\hbar = 1$ throughout):
\begin{equation}
    H = H_S + H_{SE} + H_E + H_{SB} + H_{B}~.
\end{equation}
We assume that the molecule possesses a single electronic energy level with the creation (annihilation) operator $a_0^\dagger$ ($a_0$) and energy $\varepsilon_0$:
\begin{equation}
    H_S = \varepsilon_0 \: a_0^\dagger a_0 ~.
\end{equation}
The position of the energy level in question, with respect to the Fermi energies of the unbiased electrodes, can be altered by applying the gate potential $V_\mathrm{g}$: $\varepsilon_0 = \varepsilon_{00} - \lvert e\rvert V_\mathrm{g}$ where $e$ is the electron charge.
The left (L) and right (R) electrodes are described by 
\begin{equation}
    H_E = \sum_{l=\mathrm{L},\mathrm{R}} \sum_{k_l} \epsilon_{k_l} c^\dagger_{k_l} c_{k_l} ~,
\end{equation}
where $c^\dagger_{k_l}$ ($c_{k_l}$) is the creation (annihilation) operator for the electron in the level $k_l$ with energy $\epsilon_{k_l}$ in the lead $l$.
The molecule-lead coupling is described by the usual tunnelling Hamiltonian:
\begin{equation}
    H_{SE} = \sum_{l=\mathrm{L},\mathrm{R}} \sum_{k_l} V_{k_l} a^\dagger_0  c_{k_l}  + V_{k_l}^*c^\dagger_{k_l} a_0 ~,
\end{equation}
where $V_{k_l}$ is the coupling strength. 
Furthermore, the electronic degree of freedom is coupled to a collection of thermalised vibrational modes (a phonon bath) which are modelled as harmonic oscillators with frequencies $\omega_q$ and raising (lowering) operators $b_q^\dagger$ ($b_q$):  
\begin{equation}
  H_{B} = \sum_{q} \: \omega_q b_q^\dagger b_q  ~.
\end{equation}
The electronic-vibrational coupling has the usual linear form:
\begin{equation}
  H_{SB} = \sum_{q} g_q\: a_0^\dagger a_0 (b_q^\dagger + b_q)~,
\end{equation}
where $g_q$ is the electron-vibrational coupling constant.
In general, the electronic degrees of freedom in molecular junctions can couple both to the intra-molecular as well as environmental vibrational modes, the latter originating from the solvent or the surface on which the molecule is deposited.
We shall treat these interactions on an equal footing (which is possible if one assumes thermalisation of all the molecular and broader environmental vibrational modes) through the so-called spectral density (SD).
It is formally defined as:
\begin{equation}
    \mathcal{J}(\omega) = \sum_q \lvert g_q\rvert^2 \delta(\omega - \omega_q)~,
\end{equation}
and describes the distribution of the vibrational modes weighted by the strength of the electron-vibrational coupling. 
Throughout this work we assume that all vibrational modes can be found in their thermal equilibrium state. Our approach therefore disregards any non-equilibrium vibrational dynamics although we note that it may affect the transport properties of the junction, and may be especially important in the case of intra-molecular vibrational modes.\cite{sowa2017vibrational,hartle2009vibrational} 

Let us note that the transport through molecular self-assembled monolayers\cite{fan2002charge,love2005self} can also be modelled using such Hamiltonian although the  effects of inter-molecular interactions (not included here)  may play an important role in those systems.\cite{dubi2014transport}

\section{Theory \label{GQ}}
\subsection{Polaron transformation}
We first perform the polaron (Lang-Firsov) transformation which eliminates the $H_{SB}$ term from the Hamiltonian:\cite{lang1963kinetic,mahan2013many}
\begin{equation}
\bar{H} = e^G H e^{-G}
\end{equation}
where 
%\begin{equation}
$G =\sum_{q} (g_{q}/\omega_q)\: a^\dagger_0 a_0(b^\dagger_{q} - b_{q})$.
%\end{equation}
This yields the Hamiltonian in the polaron-transformed frame:
\begin{align}
    &\bar{H} = \bar{H}_S + \bar{H}_{SE} + H_E + H_{B}~; \label{polaron}\\
    &\bar{H}_S = \left(\varepsilon_0 - \sum_q \dfrac{\lvert g_q\rvert^2}{\omega_q}\right) \: a_0^\dagger a_0 \equiv \bar{\varepsilon}_0 \: a_0^\dagger a_0 ~;\\
    &\bar{H}_{SE} = \sum_{l,k_l} V_{k_l} X^\dagger a^\dagger_0  c_{k_l}  + V_{k_l}^*c^\dagger_{k_l} X a_0  ~. \label{HSE}
\end{align}
As can be seen above, the position of the molecular electronic energy level has been renormalised. Furthermore, the displacement operators $X$ and $X^\dagger$ have been introduced into the molecule-lead coupling Hamiltonian:
\begin{equation}
X = \exp\left[-\sum_q \dfrac{g_q}{\omega_{q}}(b^\dagger_{q} - b_{q})\right].
\end{equation}
Physically, their presence in Eq.~\eqref{HSE} accounts for the fact that the charging of the molecule is accompanied by a displacement of the vibrational modes coupled to this molecular level.
Properties of the displacement operators are extensively discussed, for instance, in Ref.~\onlinecite{barnett2002methods}.

\subsection{Quantum Master Equation}
We begin with a second-order quantum master equation within the Born approximation (valid in the non-adiabatic regime\cite{nitzan2006chemical} of weak molecule-lead coupling) in the form given by Yan:\cite{yan1998quantum}
\begin{equation} \label{Born}
    \dfrac{\mathrm{d}\rho(t)}{\mathrm{d}t} = -\mathrm{i}\mathcal{L} \rho(t) - \int_0 ^t \mathrm{d}\tau \langle \mathcal{L}'(t)  \mathcal{G}(t,\tau)\mathcal{L}'(\tau)  \mathcal{G}^\dagger(t,\tau)  \rangle \rho(t) ~,
\end{equation}
where the superoperators in the above are defined as: $\mathcal{L}\bullet \equiv [\bar{H}_S,\bullet]$, $\mathcal{L'}\bullet \equiv [\bar{H}_{SE},\bullet]$, and $\langle \ldots \rangle$ denotes the trace over all (phononic and fermionic) environmental degrees of freedom. $\mathcal{G}(t,\tau)$ is the (free) system propagator: $\mathcal{G}(t,\tau) \equiv e^{-\mathrm{i}\mathcal{L}(t-\tau)}$. It can be defined in Hilbert space, as acting on an arbitrary operator $A$, by:
\begin{equation}\label{freeprop}
      \mathcal{G}(t)[A] = e^{-\mathrm{i}\bar{H}_S t} A e^{\mathrm{i}\bar{H}_S t}~.
\end{equation}
In order to go beyond the second-order Born approximation (and similarly to what is done within the self-consistent Born approximation\cite{esposito2010self,jin2014improved}) we shall replace the free system propagator $\mathcal{G}(t,\tau)$ in Eq.~\eqref{Born} with an effective one, $\mathcal{U}(t,\tau)$.
When deriving the second-order molecule-lead hopping rates, this as-yet-unknown effective propagator will account for the fact that the unitary evolution of the relevant system operators is influenced by the system-environment coupling, and allow us to incorporate the otherwise missing lifetime broadening into our description.\cite{esposito2009transport}

Let us first expand the commutators in Eq.~\eqref{Born}, and then replace the free propagator with $\mathcal{U}(t,\tau) = \mathcal{U}(t - \tau)$.
This yields:
\begin{widetext}
\begin{multline} \label{expan}
\begin{aligned}
\dfrac{\mathrm{d}\rho(t)}{\mathrm{d}t} = -\mathrm{i}[\bar{H}_S,\rho(t)] - & \sum_l\sum_{k_l}\int_0 ^t \mathrm{d}\tau \bigg\{ \bigg.\\
&   C^+_{k_l}(t-\tau) B(t-\tau) a_0 \: \mathcal{U}(t-\tau) \left[ a_0^\dagger\right] \rho(t)
- {C^-_{k_l}}^*(t-\tau) B^*(t-\tau) a_0 \: \rho(t) \mathcal{U}(t-\tau) \left[a_0^\dagger\right] \\
+ & C^-_{k_l}(t-\tau) B(t-\tau) a_0^\dagger \: \mathcal{U}(t-\tau) \left[ a_0\right] \rho(t)
- {C^+_{k_l}}^*(t-\tau) B^*(t-\tau) a_0^\dagger \: \rho(t) \mathcal{U}(t-\tau) \left[a_0\right] \\
+ & {C^-_{k_l}}^*(t-\tau)B^*(t-\tau) \:\rho(t)  \mathcal{U}(t-\tau) \left[a_0^\dagger\right] a_0 
- C^+_{k_l}(t-\tau) B(t-\tau) \: \mathcal{U}(t-\tau) \left[ a_0^\dagger\right] \rho(t)a_0  \\
+ & {C^+_{k_l}}^*(t-\tau) B^*(t-\tau) \:\rho(t) \mathcal{U}(t-\tau) \left[a_0\right]a_0^\dagger
- C^-_{k_l}(t-\tau) B(t-\tau) \: \mathcal{U}(t-\tau) \left[a_0 \right]\rho(t) a_0^\dagger \bigg.\bigg\}~,
\end{aligned}
\end{multline}
\end{widetext}
%where $\mathcal{U}(t-\tau) \left[A \right]$ denotes the propagator acting on an operator $A$.
In the equation above, we denoted
$ C^+_{k_l}(t-\tau) = \lvert V_{k_l}\rvert^2 \langle c^\dagger_{k_l}(t) c_{k_{l}}(\tau) \rangle$ and $C^-_{k_l}(t-\tau) = \lvert V_{k_l}\rvert^2 \langle c_{k_l}(t) c_{k_{l}}^\dagger(\tau) \rangle$.
$B(t-\tau)$ is the phononic correlation function defined as:
\begin{equation}
    B(t - \tau) = \langle X(t) X^\dagger (\tau)\rangle = \mathrm{Tr}\Big[X(t) X^\dagger (\tau)\rho_B\Big]~,
\end{equation}
where $\rho_B$ is the density matrix of the phonon bath. 
The fermionic correlation functions can be written as:
\begin{equation}\label{fourier}
    \sum_{k_l} C^\pm_{k_l}(t)  = \int_{-\infty}^\infty \dfrac{\mathrm{d}\epsilon}{2\pi} \ \Gamma_l^\pm (\epsilon) \ e^{\pm\mathrm{i}\epsilon t}~,
\end{equation}
where $\Gamma_l^+ (\epsilon) =  \Gamma_l(\epsilon) f_l(\epsilon)$ and
$\Gamma_l^- (\epsilon) =  \Gamma_l(\epsilon) [1 - f_l(\epsilon)]$.   
Here, $\Gamma_l(\epsilon) = 2\pi \sum_{k_l} \lvert V_{k_l}\rvert^2 \delta(\epsilon- \epsilon_{k_l}) $ is the spectral density for the lead $l$. 
The Fermi distributions in each of the leads are given by: $f_l(\epsilon) = (\exp[(\epsilon-\mu_l)/k_{\mathrm{B}} T] +1)^{-1}$, where $\mu_l$ is the chemical potential of the lead $l$.
Furthermore, for thermalised vibrational modes the phononic correlation function is given by:
\begin{multline}
B(t) = \exp \bigg[-\sum_q \dfrac{g_q^2}{\omega_q^2}\big(N_q (1 - e^{\mathrm{i}\omega_q t}) \\ + (N_q +1)(1 - e^{-\mathrm{i}\omega_q t}) \bigg) \bigg] ~,
\end{multline}\label{corr1}
where $N_q = (e^{\omega_q \beta} -1)^{-1}$ is the average excitation of the mode $q$ at the inverse temperature: $\beta = 1/k_{\mathrm{B}} T$.
Using the definition of the phononic spectral density, the above can be written in a more convenient form:
\begin{multline}\label{corr}
    B(t) = \exp\bigg[\int_0^\infty \mathrm{d}\omega \dfrac{\mathcal{J}(\omega)}{\omega^2} \bigg( \coth{\left(\dfrac{\beta\omega}{2}\right)} \times \\ \big(\cos{\omega t} - 1\big) - \mathrm{i}\sin{\omega t} \bigg)\bigg] ~.
\end{multline}

In Eq.~\eqref{expan}, we next substitute $t' = t-\tau$ and, anticipating that our interest will lie in the steady-state limit,  extend the integration limit to infinity and replace $\rho(t)$ with the stationary density matrix $\rho_{\mathrm{st}}$ such that:
$\mathrm{d}\rho_{\mathrm{st}}/\mathrm{d}t = 0$. 
The quantum master equation now takes the form:
\begin{multline}\label{SS}
0 = -\mathrm{i}\mathcal{L}\rho_{\mathrm{st}} - \sum_l 
   \int_{-\infty}^\infty \dfrac{\mathrm{d}\epsilon}{2\pi} \bigg\{ \int_0 ^\infty \mathrm{d}t'\ \Gamma_l^+ (\epsilon)  e^{+\mathrm{i}\epsilon t'} \times \\ B(t') a_0 \: \mathcal{U}(t') \left[ a_0^\dagger\right] \rho_{\mathrm{st}}
- ... \bigg\} ~,
\end{multline}
and similarly for the rest of the terms.
The solution to the above QME is the stationary density matrix, $\rho_{\mathrm{st}}$, which will later be used to calculate the current flowing through the junction. 

\subsection{The effective propagator}
Before we determine the effective propagator required in Eq.~\eqref{SS}, let us briefly discuss the free propagator present in Eq.~\eqref{Born}. It is defined in Eq.~\eqref{freeprop}, and in the present case it yields: $\mathcal{G}(t)[a_0] = a_0 e^{\mathrm{i} \bar{\varepsilon}_0 t}$, and $\mathcal{G}(t)[a_0^\dagger] = a_0^\dagger e^{-\mathrm{i} \bar{\varepsilon}_0 t}$. 
One can easily see the origin of the problem encountered within the standard Born approximation. Since, in the second-order QME \eqref{Born}, it is assumed that the evolution of the creation (annihilation) operator is unaffected by the molecule-lead coupling, the electron hopping described by this dissipator will not include lifetime broadening.
Therefore, we replace the free propagator with:
\begin{equation}
    \mathcal{U}(t)[A] = \mathrm{Tr}\left[e^{-\mathrm{i}\bar{H} t} A e^{\mathrm{i}\bar{H} t} \right]~,
\end{equation}
where $A = \{a_0, a_0^\dagger \}$, \textit{c.f.} Ref.~\onlinecite{esposito2009transport}.
In order to determine $\mathcal{U}(t)[a_0]$, let us consider the equations of motion for $a_0(t) \equiv \exp\left(-{\mathrm{i}\bar{H} t}\right) a_0 \exp\left({\mathrm{i}\bar{H} t}\right)$ and $c_{k_l}(t) \equiv \exp\left(-{\mathrm{i}\bar{H} t}\right) c_{k_l} \exp\left({\mathrm{i}\bar{H} t}\right)$ operators in the polaron-transformed frame:
\begin{align}
&\dot{a}_0(t) = \mathrm{i} \bar{\varepsilon}_0 a_0(t) - \mathrm{i} \ \sum_{l,k_l} V_{k_l} X^\dagger(t) c_{k_l}(t)   ~;\\
&\dot{c}_{k_l}(t) = \mathrm{i} \epsilon_{k_l} c_{k_l}(t) - \mathrm{i} \ V^*_{k_l} X(t) a_0(t) ~.
\end{align}
Using the Laplace transform, $a_0(z) = \int_0^\infty \mathrm{d} t e^{-z t} a_0(t)$, we can turn this set of differential equations into an  algebraic one:
\begin{align}
& z a_0(z) - a_0(0) = \mathrm{i} \bar{\varepsilon}_0 a_0(z) - \mathrm{i}  \sum_{l,k_l} V_{k_l} X^\dagger(z) c_{k_l}(z)   ~;\\
& z c_{k_l}(z) - c_{k_l}(0) = \mathrm{i} \epsilon_{k_l} c_{k_l}(z) - \mathrm{i} \ V^*_{k_l} X(z) a_0(z) ~.
\end{align}
Eliminating the fermionic reservoir modes gives
\begin{multline}\label{laplace2}
 [z - \mathrm{i}\bar{\varepsilon}_0 ] \  a_0(z) =  a_0(0) -  \sum_{l,k_l} \dfrac{\lvert V_{k_l}\rvert^2}{z - \mathrm{i} \epsilon_{k_l}} a_0 (z)\\  - \mathrm{i} \sum_{l,k_l} \dfrac{V_{k_l} X^\dagger(z) }{z -\mathrm{i} \epsilon_{k_l}} c_{k_l} (0)~,   
\end{multline}
where, crucially, the displacement operators have cancelled in the second term on the right-hand-side. 
We disregard the final term (which vanishes when tracing out the fermionic reservoirs), and replace the sum in the second term with an integral:
\begin{equation}
    \sum_{k_l} \dfrac{\lvert V_{k_l}\rvert^2}{z -\mathrm{i} \epsilon_{k_l}} \ \rightarrow \ \int_{-\infty}^\infty \dfrac{\mathrm{d}\epsilon_l}{2\pi} \ \dfrac{\Gamma_l(\epsilon_l)}{z - \mathrm{i}\epsilon_l} ~.
\end{equation}
We will assume that $\Gamma_l(\epsilon_l)$ is a Lorentzian,
$   \Gamma_l (\epsilon_l) = \dfrac{\Gamma_l \  \delta_l^2}{\epsilon_l^2 + \delta_l^2}~, $
and to obtain the wide-band approximation, set $\delta_l \rightarrow \infty$.\cite{malz2018current} 
Consequently,
\begin{equation}
a_0(z) = \dfrac{1}{z - \mathrm{i}\bar{\varepsilon}_0 + (\Gamma_\mathrm{L} + \Gamma_\mathrm{R})/2 } \ a_0(0) ~.
\end{equation}
Moving back to the time domain, and defining $\Gamma = \left(\Gamma_\mathrm{L} + \Gamma_\mathrm{R} \right)/2$, yields:
\begin{equation}\label{SCC1}
\mathcal{U}(t) \left[a_0\right] = a_0 e^{+\mathrm{i}\bar{\varepsilon}_0 t  - \Gamma t }~,   
\end{equation}
and similarly, for $a_0^\dagger$, we obtain:
\begin{equation}\label{SCC2}
\mathcal{U}(t) \left[a_0^\dagger\right] = a_0^\dagger e^{-\mathrm{i}\bar{\varepsilon}_0  - \Gamma t }~.    
\end{equation}
Let us stress here that the simplicity of the correction to the Born approximation ($e^{-\Gamma t}$) is only possible due to the (not immediately obvious) cancellation of the displacement operators in Eq.~\eqref{laplace2}.
We note that the effective propagator of the type introduced here may overestimate the amount of lifetime broadening, see Ref.~\onlinecite{flensberg2003tunneling} for a detailed discussion. While more sophisticated approaches to determining the effective evolution in this context have been developed (such as the self-consistent approach of Galperin \textit{et al.}\cite{galperin2006resonant}), their complexity would prevent us from obtaining closed-form expressions for the electric current through the junction -- one of the main objectives of this work. 

\subsection{Back to the Quantum Master Equation}
Once again, we will take $V_{k_l}=V_l =$ const. and assume the wide-band approximation. Then, $\Gamma_l(\epsilon)$ in Eq.~\eqref{fourier} becomes $\Gamma_l = 2\pi \lvert V_{l}\rvert^2 \varrho_l $ where $\varrho_l$ is the constant density of states in lead $l$.
Inserting Eqs.~(\ref{SCC1}, \ref{SCC2}) into the Quantum Master Equation, allows us to express it in a surprisingly simple form:
\begin{multline} \label{Lind}
0 = -\mathrm{i}[\bar{H}_S,\rho_{\mathrm{st}}] + \sum_l \bigg\{ \upsilon_l \left(a_0^\dagger \rho_{\mathrm{st}} a_0 - a_0 a_0^\dagger \rho_{\mathrm{st}}\right) \\
+ \upsilon^*_l \left(a_0^\dagger \rho_{\mathrm{st}} a_0 - \rho_{\mathrm{st}} a_0 a_0^\dagger \right)
+ \bar{\upsilon}_l \left(a_0 \rho_{\mathrm{st}} a_0^\dagger -  a_0^\dagger a_0 \rho_{\mathrm{st}}\right) \\
+ \bar{\upsilon}^*_l \left(a_0 \rho_{\mathrm{st}} a_0^\dagger - \rho_{\mathrm{st}} a_0^\dagger a_0  \right)  \bigg\}~,
\end{multline}
where the rates $\upsilon_l$ and $\bar{\upsilon}_l$ can be inferred from Eq.~\eqref{SS}.
The solution to the above, in the basis of the neutral and charged molecular states, can be written as:
% $\{ \lvert N\rangle, \lvert N+1\rangle\}$
\begin{equation}\label{SSS}
    \rho_{\mathrm{st}} = \Bigg(\begin{array}{c c}
  \dfrac{\bar{\gamma}_L +\bar{\gamma}_R}{\gamma_\mathrm{L} + \gamma_\mathrm{R} + \bar{\gamma}_L +\bar{\gamma}_R} & 0 \\ 
  0 &   \dfrac{\gamma_\mathrm{L} + \gamma_\mathrm{R}}{\gamma_\mathrm{L} + \gamma_\mathrm{R} + \bar{\gamma}_L +\bar{\gamma}_R}
 \end{array} \Bigg) ~,
\end{equation}
where the rates in the above are defined  as $\gamma_l = 2\mathrm{Re}[\upsilon_l]$, $\bar{\gamma}_l = 2\mathrm{Re}[\bar{\upsilon}_l]$, and given by:
\begin{widetext}
\begin{align}\label{R1}
& \gamma_l =  2\ \mathrm{Re} \Bigg[ \Gamma_l \int_{-\infty}^\infty \dfrac{\mathrm{d}\epsilon}{2\pi} f_l(\epsilon) \int_0^\infty \mathrm{d}\tau \ e^{+\mathrm{i}(\epsilon - \bar{\varepsilon}_0 )\tau} e^{-\Gamma \tau} B(\tau)\Bigg] ~;\\     
& \bar{\gamma}_l = 2\ \mathrm{Re} \Bigg[ \Gamma_l \int_{-\infty}^\infty \dfrac{\mathrm{d}\epsilon}{2\pi} [1 - f_l(\epsilon)] \int_0^\infty \mathrm{d}\tau \ e^{-\mathrm{i}(\epsilon - \bar{\varepsilon}_0 )\tau} e^{-\Gamma \tau} B(\tau)\Bigg] ~. \label{R2}
\end{align}
\end{widetext}
We note that coherences between the two charge states vanish (and are generally decoupled from the electronic populations). This is in fact universally true, and is not a consequence of the approximations made herein.\cite{leijnse2008kinetic} The quantum master equation \eqref{Lind} has a secular form, and consequently we only need to consider the real parts of the response functions, as shown above. 
\begin{figure*}
    \centering
    \includegraphics{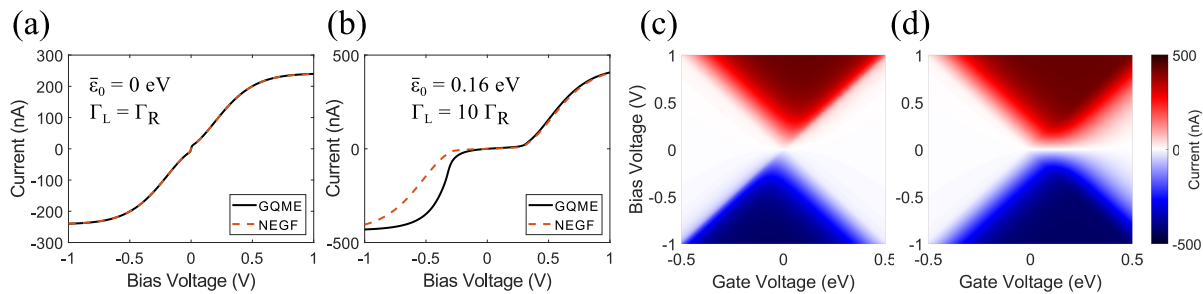}
    \caption{Transport characteristics for an electronic level coupled to a super-ohmic phonon bath with $\lambda = 150$ meV and $\Omega_c = 25$ meV. (a) $IV$ characteristics on resonance: $\bar{\varepsilon}_0 = 0$ meV. (b) $IV$ characteristics off resonance with $\bar{\varepsilon}_0 = 160$ meV. (c,d) Stability diagrams calculated for the case asymmetric coupling $\Gamma_\mathrm{L} = 10 \Gamma_\mathrm{R}$ using the (c) GQME, and (d) NEGF approach. $T=10$ K and $\Gamma_\mathrm{R} = 2$ meV throughout.}
    \label{F1}
\end{figure*}
The steady-state current through the molecule at the left and right contacts is equal and opposite. Let us consider the right contact where it can be expressed (again, in the units of $\hbar$) simply as:
%\begin{equation}
$    I = e \left(\ \gamma_\mathrm{R} \rho_{00, \mathrm{st}} - \bar{\gamma}_R \rho_{11, \mathrm{st}} \right)$.\cite{sowa2017environment,flindt2004full}
Together with Eq. \eqref{SSS} this gives the generic expression:
\begin{equation} \label{current}
    I = e \ \dfrac{\gamma_\mathrm{L} \bar{\gamma}_R - \gamma_\mathrm{R} \bar{\gamma}_L}{\gamma_\mathrm{L} + \gamma_\mathrm{R} + \bar{\gamma}_L +\bar{\gamma}_R} ~.
\end{equation}
The above expression, together with Eqs.~(\ref{R1},\ref{R2}), constitutes the GQME approach as denoted in Fig.~\ref{scheme1}, and will serve as a basis for the remainder of this work.

\subsection{Comparison to NEGF \label{comparison}}
Firstly, let us note that in the limit of vanishing electron-phonon coupling [when $B(t) = 1$], the above approach recovers the usual Landauer-B\"{u}ttiker (LB) expression for current through a single non-interacting level, see Appendix A.

In the presence of electron-phonon coupling, the expression \eqref{current} together with Eqs. (\ref{R1}, \ref{R2})  yields a result that is encouragingly similar to that derived using the non-equilibrium Green function (NEGF) approach,\cite{wingreen1989inelastic,jauho1994time,glazman1988inelastic} for convenience given in Appendix B. The difference between the NEGF result given in Eq.~\eqref{NEGF1} and the one derived here stems predominantly from the fact that our result was derived by first determining the steady-state density matrix $\rho_{\mathrm{st}}$, and consequently it accounts for the charge state of the molecular system [unlike Eq.~\eqref{NEGF1}, see Ref.~\onlinecite{wingreen1989inelastic}].
This is perhaps best demonstrated with a numerical example.
In what follows, we apply the bias voltage symmetrically $\mu_l = \pm e V_\mathrm{b}/2$ and, for simplicity, assume that the electronic level considered here is coupled to a phonon bath with a structureless super-ohmic spectral density:
\begin{equation}\label{SD}
    \mathcal{J}(\omega) = \dfrac{\lambda}{2} \left(\dfrac{\omega}{\Omega_{c}}\right)^3 e^{-\omega/\Omega_c}~,
\end{equation}
where $\lambda$ is the reorganisation energy, and $\Omega_c$ is the cut-off frequency.

First, in Fig.~\ref{F1}(a) we show the $IV$ characteristics calculated for the case of symmetric molecule-lead coupling ($\Gamma_\mathrm{L} = \Gamma_\mathrm{R}$) and the molecular energy level being on resonance ($\bar{\varepsilon}_0=0$). 
The GQME and NEGF results give rise to identical $IV$ characteristics (it can be shown that this is always the case on resonance, i.e. when $\bar{\varepsilon}_0=0$).

The more interesting case is one of asymmetric molecule-lead coupling ($\Gamma_\mathrm{L} \ne \Gamma_\mathrm{R}$) when the molecule does not lie on resonance ($\bar{\varepsilon}_0 \ne 0$). If one of the rates is much greater than the other, the molecule is almost always occupied (or empty, depending on the sign of $V_\mathrm{b}$). That means that depending on the direction of the current it is either the hopping on or off the molecule that is the overall bottleneck of the transport. In the presence of vibrational coupling and when the molecular level lies off resonance, this means that different inelastic processes control the overall current which results in current rectification (asymmetric $IV$ characteristics).\cite{hartle2011resonant}
This effect has been observed experimentally,\cite{lau2015redox,wu2004control} and as shown in Fig.~\ref{F1}(b) is captured by the GQME approach. On the other hand, the NEGF result always remains symmetric with respect to the direction of the current flow as can be easily seen from Eq.~\eqref{NEGF1}  (as long as the bias is applied symmetrically). 
In Figs.~\ref{F1}(c,d) we show the stability diagrams (maps of current as a function of the applied bias and gate voltage) calculated using the NEGF and GQME approaches for the case of asymmetric molecule-lead coupling. 
The GQME quantum master equation once again captures the asymmetry of the transport characteristics whereas the NEGF approach predicts a result that is symmetric with respect to the bias voltage. 
More sophisticated NEGF approaches (of the type first developed by Galperin \textit{et al.}\cite{galperin2006resonant}) can capture the current rectification pictured in Figs.~\ref{F1}(b,c)\cite{volkovich2011bias} but they do not posses the appealing simplicity of the methods considered herein.

In summary, our result is in a complete agreement with the scattering approach given in Appendix \ref{AppB} on resonance, as well as for vanishing electron-phonon coupling. In the case of asymmetric molecule-lead coupling, it can also account for current rectification (which as discussed above is not simply an artefact of the GQME method) -- an effect not captured by the scattering approach considered here. 
\begin{figure*}
    \centering
    \includegraphics{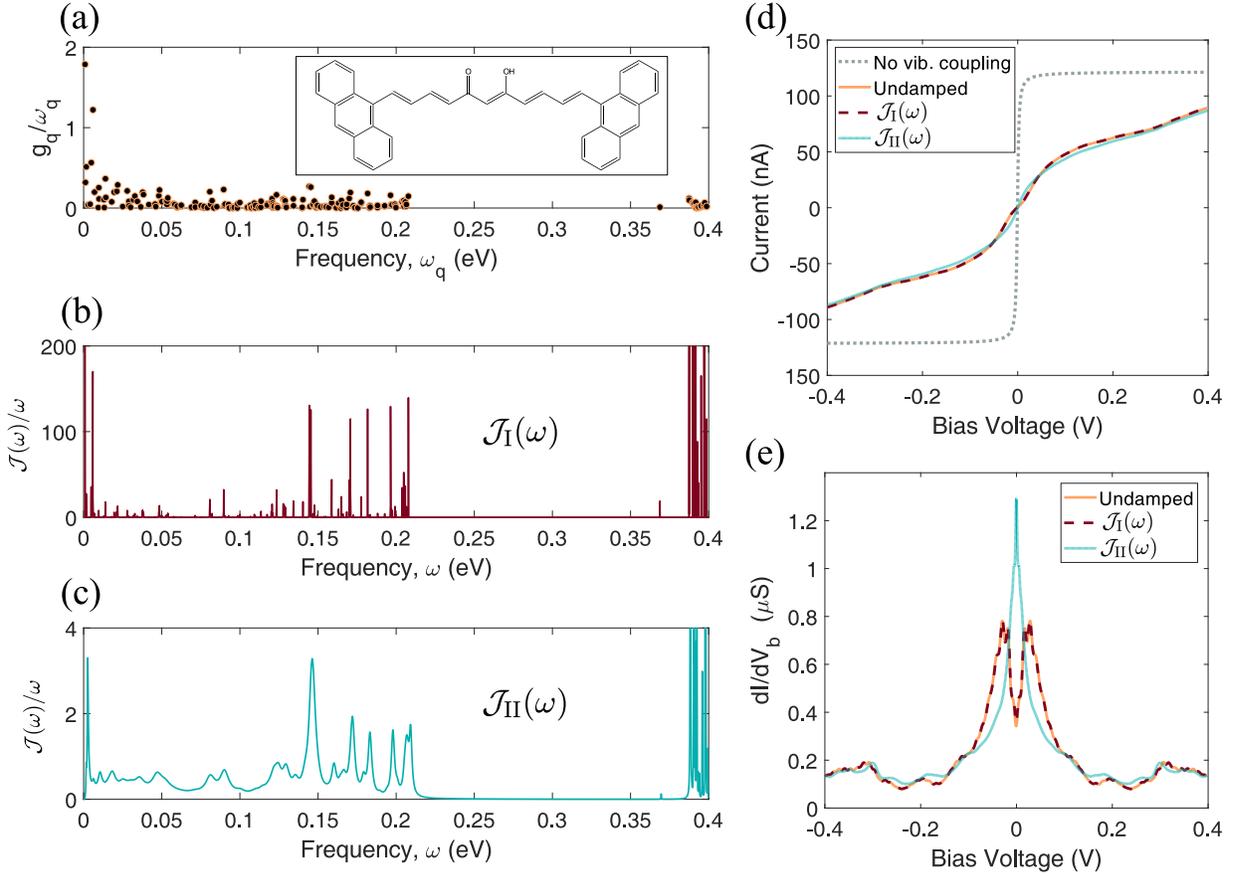}
    \caption{(a) Dimensionless electron-phonon coupling parameters as a function of the vibrational frequencies $\omega_q$, from Ref.~\onlinecite{burzuri2016sequential}. Inset: the molecular structure of the curcuminoid molecule considered therein. (b,c) Continuous spectral densities obtained from the above coupling parameters with (b) $\eta = 1$ meV, $\Lambda = 50$ meV, and (c) $\eta = 50$ meV, $\Lambda = 60$ meV. (d) $IV$ characteristics, and (e) their derivatives (differential conductance; $\mathrm{d}I/\mathrm{d}V_\mathrm{b}$) obtained for transport through the 9ALCccmoid-based junction. $\Gamma_\mathrm{L} = \Gamma_\mathrm{R} = 1$ meV, $\bar{\varepsilon}_0 = 0$, and $T=4$ K in both instances.}
    \label{9ALC1}
\end{figure*}

\subsection{Microscopically-motivated coupling \label{DFT}}
Before we proceed to study the transport behaviour at higher temperatures, we first take this opportunity and use our GQME result to analyse the low-temperature transport characteristics of a particular molecular system with the aid of DFT calculations.
We shall consider a molecular junction incorporating the curcuminoid-based molecule, Fig.~\ref{9ALC1}(a), recently studied experimentally by Burzur{\'\i} and coworkers.\cite{burzuri2016sequential} 
The authors estimated the electron-vibrational coupling strength (of the anionic charge state) for each of the molecular vibrational modes using the method described in Ref.~\onlinecite{seldenthuis2008vibrational}, as follows: Firstly, one performs the geometry optimisation of the neutral and charged (here anionic) structure. By comparing the two geometries the so-called Duschinsky shift vector can be obtained from which the dimensionless $g_q/\omega_q$ quantity can be easily determined for each of the molecular vibrational modes.\cite{barone2009fully,duschinsky1937importance} The result of this calculation (from Ref.~\onlinecite{burzuri2016sequential}) is for convenience shown in Fig.~\ref{9ALC1}(a).

At this point there are two ways in which we can proceed.
We can either (i) assume the molecular vibrational modes are infinitely long-lived, or (ii) introduce a small amount of damping and combine all these electron-vibrational couplings into a single smooth spectral density. 
In both instances the total spectral density can be obtained simply as a sum:
\begin{equation}
    \mathcal{J}_{\mathrm{tot}}(\omega) = \sum_q \mathcal{J}_{q}(\omega)~,
\end{equation}
where $\mathcal{J}_{q}(\omega)$ is a spectral density for the vibrational mode $q$. For the the infinitely long-lived (undamped) modes, it is simply: $\mathcal{J}_{q}(\omega) = \lvert g_q\rvert^2 \delta(\omega - \omega_q)$.
In the latter case, we shall assume that $\mathcal{J}_{q}(\omega)$ takes the form given by Roden \textit{et al.}:\cite{roden2012accounting}
\begin{multline}
    \mathcal{J}_{q}(\omega) =\\  \dfrac{(g_q/\omega_q)^2 \ \eta \ \omega^3 \ e^{-\omega/\Lambda} \Theta(\omega)}{\pi^2 \eta^2 \omega^2 e^{-2\omega/\Lambda} + (\omega -\omega_q + \eta\Lambda -\eta \omega e^{-\omega/\Lambda} \mathrm{Ei}(\omega/\Lambda) )^2}~,
\end{multline}
where $\Theta(\omega)$ is the unit step function, and $\mathrm{Ei}(x)$ denotes the exponential integral.
The above expression was obtained by assuming that the molecular modes are coupled to a secondary ohmic phonon bath (with a cut-off frequency $\Lambda$ and a reorganisation energy $\eta \Lambda$).
Figs.~\ref{9ALC1}(b,c) show two damped spectral densities obtained in this way from the \textit{ab initio} vibrational coupling strengths plotted in Fig.~\ref{9ALC1}(a) for illustrative values of $\eta$ and $\Lambda$.

We begin, in Fig.~\ref{9ALC1}(d), by calculating the $IV$ characteristics on resonance at $T = 4$ K for the case of coupling to the undamped vibrational modes, the continuous spectral densities from Figs.~\ref{9ALC1}(b,c), as  well as the case of no vibrational coupling (the Landauer-B\"{u}ttiker limit).
Coupling to undamped and only slightly damped [$\mathcal{J}_{\mathrm{I}}(\omega)$, Fig.~\ref{9ALC1}(b)] vibrational modes reassuringly delivers almost identical behaviour.
Significantly different characteristics can be observed in the case of much greater damping [$\mathcal{J}_{\mathrm{II}}(\omega)$, Fig.~\ref{9ALC1}(c)].
These differences are clearly visible in the plot of differential conductance ($\mathrm{d}I/\mathrm{d}V_\mathrm{b}$) in Fig.~\ref{9ALC1}(e). At zero bias, the junction is markedly less conductive in the case of coupling to the undamped molecular modes. This can be explained by significant damping of the low-frequency vibrational modes in $\mathcal{J}_{\mathrm{II}}(\omega)$.

Somewhat surprisingly, the structure of the transport characteristics [peaks in the differential conductance, Fig.~\ref{9ALC1}(e)] even in the undamped case cannot be easily correlated with the individual molecular vibrational modes. The origin of this behaviour can be traced back to the two low-frequency vibrational modes that are strongly coupled (with $g_q/\omega_q > 1$) to the electronic degree of freedom, see Fig.~\ref{9ALC1}(a).
To understand this effect let us consider transport through an electronic level coupled to a vibrational mode with frequency $\omega_Q$ (with an intermediate coupling strength, $g_q/\omega_q < 1$).
In the absence of strong coupling to any low-frequency modes, one would expect to observe a peak in the differential conductance for bias voltage such that $e V_\mathrm{b} = 2 \omega_Q$ (for symmetrically applied bias, and on resonance).
This  corresponds to an excitation of the vibrational mode in question, and a simultaneous ground-state-to-ground-state transition for all the remaining vibrational modes.
However, when the molecular electronic level is also strongly coupled to a certain low-frequency mode (with frequency $\omega_P$), the aforementioned transition is no longer efficient due to the poor Franck-Condon overlap for the ground-state-to-ground-state transition of the mode $P$.
Instead, the visible peaks in the differential conductance (corresponding to efficient inelastic transitions) occur for the simultaneous excitation of both the $Q$ and $P$ modes, and therefore at $e V_\mathrm{b} = 2 (\omega_Q + n \omega_P)$ where $n$ is a positive integer.
The situation complicates further when, as is the case here, $\omega_P$ is smaller than the lifetime broadening $\Gamma$.
Then, the off-set of the original peak is no longer simply an integer amount of $\omega_P$ since the individual excitations of mode $P$ can no longer be resolved.
\begin{figure}
    \centering
    \includegraphics{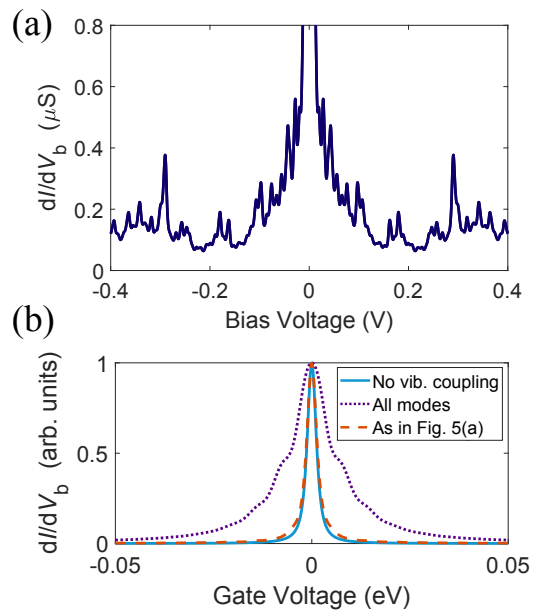}
    \caption{(a) Differential conductance on resonance ($\bar{\varepsilon}_0 = 0$) assuming coupling to the undamped modes in Fig.~\ref{9ALC1}(a) with the exception of the two strongly-coupled, low-frequency modes. (b) Renormalised zero-bias conductance as a function of the gate voltage in the case of coupling to undamped vibrational modes: $\bar{\varepsilon}_0 = - \lvert e \rvert V_\mathrm{g}$. Molecule-lead coupling - $\Gamma_\mathrm{L} = \Gamma_\mathrm{R} = 1$ meV; $T = 4$ K.}
    \label{9ALC2}
\end{figure}
Fig.~\ref{9ALC2}(a) shows the differential conductance calculated for parameters as in Fig.~\ref{9ALC1}(e) but ignoring the two strongly coupled low-frequency modes. Disregarding coupling to these modes recovers the expected behaviour -- the differential conductance now features a set of peaks which can easily be correlated with specific vibrational excitations. 

We also demonstrate the effect of these low-frequency modes on the zero-bias conductance. Fig.~\ref{9ALC2}(b) shows $\mathrm{d}I(V_\mathrm{b} = 0)/\mathrm{d}V_\mathrm{b}$ as a function of the gate voltage. The coupling to the low-frequency modes significantly affects the width of the conductance peak, which is often used to extract the lifetime broadening $\Gamma$. The values extracted in such a way should therefore be treated with caution.\footnote{Throughout this Section, we have used a significantly smaller $\Gamma$ (1 meV) than extracted in Ref.~\onlinecite{burzuri2016sequential}. Using the latter value (10 meV) washes away the vibrational features of the kind reported in Ref.~\onlinecite{burzuri2016sequential}.}

Finally, we once again stress that our approach assumes that (here, the intra-molecular) vibrational modes are thermalised at all times (to the same temperature as the fermionic reservoirs). This constitutes an important limitation of our approach, since the above assumption is not necessarily always valid as it has been observed experimentally\cite{lau2015redox}, and discussed theoretically.\cite{gelbwaser2018high,sowa2018spiro,hartle2009vibrational,hartle2011vibrational,koch2005franck}

\section{High(er)-Temperature Limits \label{highT}}
In this Section, we derive a number of approximations of Eqs.~(\ref{R1}, \ref{R2}) valid at increasing temperatures.
This collection of results yields a tiered set of approximations to the original GQME, schematically pictured in Fig.~\ref{scheme1}. 
As we shall demonstrate, two different approximations need to be made within the GQME to arrive at the well known Marcus theory -- one regarding the phononic correlation function, and one regarding the $e^{-\Gamma \tau}$ correction. 
The order in which they ought to be made (and thus the pathway taken in Fig.~\ref{scheme1}) depends on the relative magnitudes of the lifetime broadening (quantified by $\Gamma$) and the cut-off frequency of the phonon bath (describing the distribution of the vibrational modes).

\subsection{Born-Markov approximation \label{BMapp}}
We begin by recognising that the  $e^{-\Gamma \tau}$ term only leads to broadening of the $IV$ characteristics. Broadening of the Fermi distributions in the leads (at non-zero $T$) will have the same effect, and so the aforementioned correction can be ignored in the limit of $k_{\mathrm{B}}T \gg \Gamma$. This trivially recovers the result obtained within the usual second order Born-Markov (BM) approximation with respect to the leads (in the polaron-transformed frame):\cite{breuer2002theory,sowa2017environment}
%\begin{widetext}
\begin{align}\label{BM1}
& \gamma_l^{\mathrm{BM}} =  2\ \mathrm{Re} \Bigg[ \Gamma_l \int_{-\infty}^\infty \dfrac{\mathrm{d}\epsilon}{2\pi} f_l(\epsilon) \int_0^\infty \mathrm{d}\tau \ e^{+\mathrm{i}(\epsilon - \bar{\varepsilon}_0 )\tau} B(\tau)\Bigg] ~;\\     
& \bar{\gamma}_l^{\mathrm{BM}} = 2\ \mathrm{Re} \Bigg[ \Gamma_l \int_{-\infty}^\infty \dfrac{\mathrm{d}\epsilon}{2\pi} [1 - f_l(\epsilon)] \int_0^\infty \mathrm{d}\tau \ e^{-\mathrm{i}(\epsilon - \bar{\varepsilon}_0 )\tau}  B(\tau)\Bigg] ~. \label{BM2}
\end{align}
%\end{widetext}
%This perturbative result is applicable in the case of molecular junctions with very weak molecule-lead coupling such as those based on molecules connected to graphene nano-junctions via $\pi$-$\pi$ stacking.

\subsection{Marcus theory \label{MarcusT}}
In order to derive the semi-classical Marcus theory (MT), we take the limit of high temperature in Eqs.~(\ref{BM1},\ref{BM2}) and assume a slowly fluctuating (low-frequency) environment.\cite{may2008charge}
For the phononic correlation function, Eq.~\eqref{corr}, this allows us to approximate: $\coth(\beta \omega/2) \approx {2/\beta\omega}$, $\cos(\omega t) -1 \approx - \omega^2 t^2/2$, and $\sin(\omega t)\approx \omega t$. This simplifies the phononic correlation function to
\begin{equation}\label{corr2}
    B(t) \approx \exp\left(-\lambda t^2/\beta - \mathrm{i} \lambda t \right) ~,
\end{equation}
where $\lambda$ is the reorganisation energy, formally defined as:
\begin{equation}
\lambda = \int_0^\infty \mathrm{d}\omega \ \dfrac{\mathcal{J}(\omega)}{\omega} = \sum_q \dfrac{\lvert g_q\rvert^2}{\omega_q} ~. 
\end{equation}
Performing the one-sided Fourier transform leads to:
\begin{multline}
\gamma_l^{\mathrm{MT}} =  2\ \Gamma_l \int_{-\infty}^\infty \dfrac{\mathrm{d}\epsilon}{2\pi} f_l(\epsilon) \sqrt{\dfrac{\pi}{4\lambda k_{\mathrm{B}} T}} \\ \times \exp\left( -\dfrac{[\lambda-(\epsilon - \bar{\varepsilon}_0)]^2}{4\lambda k_{\mathrm{B}} T}\right) ~;   
\end{multline}
\begin{multline}
\bar{\gamma}_l^{\mathrm{MT}} = 2\ \Gamma_l \int_{-\infty}^\infty \dfrac{\mathrm{d}\epsilon}{2\pi} [1 - f_l(\epsilon)] \sqrt{\dfrac{\pi}{4\lambda k_{\mathrm{B}} T}} \\ \times \exp\left( -\dfrac{[\lambda + (\epsilon - \bar{\varepsilon}_0)]^2}{4\lambda k_{\mathrm{B}} T}\right) ~.
\end{multline} 
These are the well-known expressions for the rates of the electron hopping between a metallic electrode and a molecular energy level, as described by Marcus theory\cite{marcus1985electron,marcus1956theory,nitzan2006chemical} (in the field of electrochemistry it sometimes referred to as Marcus-DOS or Marcus-Hush-Chidsey theory).\cite{chidsey1991free,zeng2014simple,finklea2001effect,henstridge2012marcus}
These expressions have been commonly used to study the charge transport through redox molecular junctions\cite{jia2016covalently,migliore2013irreversibility,migliore2012relationship,kuznetsov2008coulomb,kuznetsov2009coulomb,yuan2018transition,bueno2016mesoscopic} also in the case of multiple transport channels.\cite{migliore2011nonlinear}

\subsection{Modified Marcus theory\label{IMarcusT}}
A major advantage of Marcus theory, besides its apparent simplicity, is the fact that the entire complexity of the phononic spectral density (which is often unknown in molecular systems) is reduced to a single parameter: the reorganisation energy, $\lambda$. This is especially important in the context of experimental studies where $\lambda$ can act as a fitting parameter.
However, as we shall  demonstrate (also see Ref.~\onlinecite{thomas2018}), MT often inadequately describes the low-temperature behaviour of molecular junctions: it severely overestimates vibrational effects at low frequencies, and underestimates them at high frequencies [which is a  consequence of the approximations made to arrive at Eq.~\eqref{corr2}].
Our goal here is to develop an extension of Marcus theory which will (at least partially) rectify those shortcomings at intermediate temperatures, whilst retaining most of its simplicity.
\begin{figure*}
    \centering
    \includegraphics{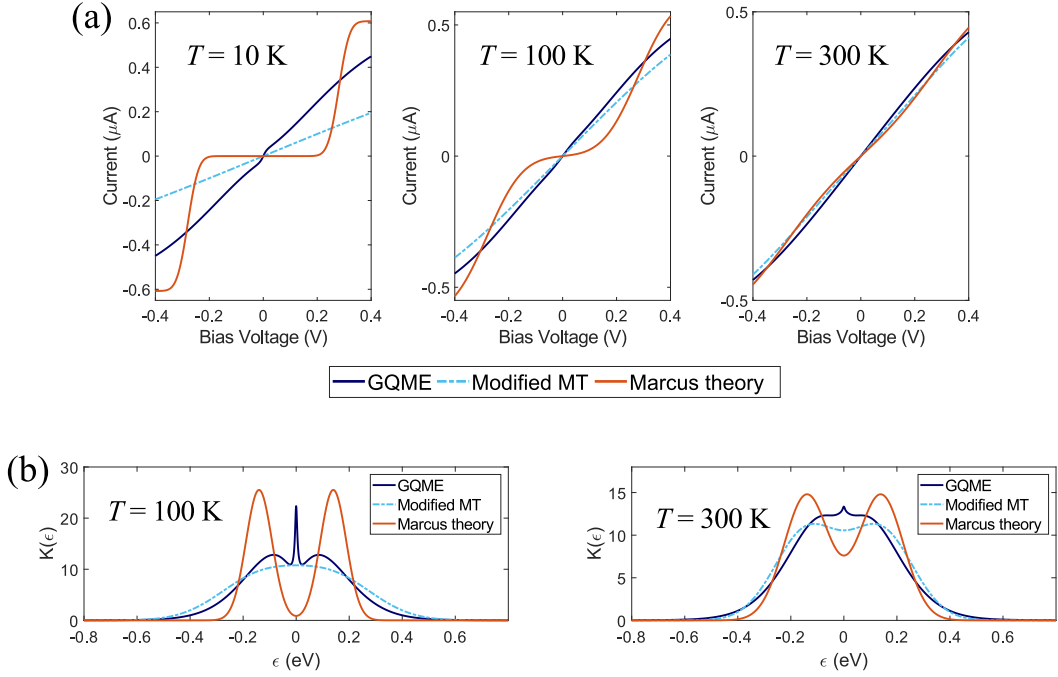}
    \caption{(a) $IV$ characteristics on resonance ($\bar{\varepsilon}_0 = 0$ eV) at different temperatures. The molecular energy level is coupled to a vibrational environment with the SD of Eq.~\eqref{SD} with $\lambda = 140$ meV and $\Omega_c = 25$ meV ($\chi = 0.0011\ \mathrm{eV}^3$); $\Gamma_\mathrm{L} = \Gamma_\mathrm{R} = 5$ meV.
    (b) Plots of $K(\epsilon)$, sum of the energy-dependent hopping rates, for parameters as in (a).}
    \label{Temp_one}
\end{figure*}

In deriving Marcus theory, we have approximated $\coth(\beta \omega/2)$ as ${2/\beta\omega}$. Here we shall expand this term to the second order: $\coth\left(\beta \omega/2 \right) \approx 2/\beta \omega + \beta \omega/6$.
Following the same procedure as in the previous section yields an almost equally simple result:
\begin{widetext}
\begin{align}\label{IMT1}
& \gamma_l^{\mathrm{MMT}} =  2\ \Gamma_l \int_{-\infty}^\infty \dfrac{\mathrm{d}\epsilon}{2\pi} f_l(\epsilon) \sqrt{\dfrac{\pi}{4\lambda k_{\mathrm{B}} T + \chi/3k_{\mathrm{B}} T}} \exp\left( -\dfrac{[\lambda-(\epsilon - \bar{\varepsilon}_0)]^2}{4\lambda k_{\mathrm{B}} T + \chi/3k_{\mathrm{B}} T}\right) ~;\\     
& \bar{\gamma}_l^{\mathrm{MMT}} = 2\ \Gamma_l \int_{-\infty}^\infty \dfrac{\mathrm{d}\epsilon}{2\pi} [1 - f_l(\epsilon)] \sqrt{\dfrac{\pi}{4\lambda k_{\mathrm{B}} T + \chi/3k_{\mathrm{B}} T}} \exp\left( -\dfrac{[\lambda+(\epsilon - \bar{\varepsilon}_0)]^2}{4\lambda k_{\mathrm{B}} T + \chi/3k_{\mathrm{B}} T}\right) ~.\label{IMT2}
\end{align}  
\end{widetext}
In the above, a new parameter describing the electron-phonon coupling ($\chi$) has been introduced
\begin{equation}
    \chi = \int_0^\infty \mathrm{d} \omega\ \omega \times \mathcal{J}(\omega) = \sum_q {\lvert g_q\rvert^2}{\omega_q}~.
\end{equation}
The rates in Eqs.~(\ref{IMT1},\ref{IMT2}) constitute what we refer to here as the \textit{Modified Marcus theory}. It is clear that  $\chi/3k_{\mathrm{B}} T$ acts as a low-temperature correction to MT which vanishes for high $T$.
As we shall demonstrate (\textit{vide infra}), the Modified Marcus theory removes some of the artefacts of MT while largely retaining its mathematical simplicity.

What is the physical meaning of the parameter $\chi$? It accounts for the coupling to higher frequency vibrational modes for which the high-temperature assumption of MT is not justified. The correction in Eqs.~(\ref{IMT1},\ref{IMT2}) thus vanishes if the vibrational modes have frequencies much lower than $k_{\mathrm{B}} T$.
It is also clear that the hopping rates derived above do not have the typical Arrhenius form (which is however recovered at high temperatures).

Let us now demonstrate the effectiveness of the Modified Marcus theory on a numerical example. 
We will again assume that the electron-phonon coupling can be described by the spectral density given in Eq.~\eqref{SD} and consider a case of coupling to relatively high-frequency vibrational modes.
In Fig.~\ref{Temp_one}, we plot the $IV$ characteristics obtained using various theoretical approaches for increasing temperature (left-hand-side of diagram in Fig.~\ref{scheme1}). Due to relatively small lifetime broadening, the GQME and BM approaches yield very similar results, and we shall omit the latter for clarity. Marcus approach gives rise to certain artefacts at lower temperatures: current suppression at low bias voltage (effectively a spurious Franck--Condon blockade), and current plateaus at artificially low $V_\mathrm{b}$. The origin of these features can be directly traced back to the assumptions made in the derivation of Marcus theory (Section \ref{MarcusT}). As can be seen in Fig.~\ref{Temp_one}, the Modified Marcus theory successfully rectifies these shortcomings at intermediate temperatures (at least at a qualitative level). 
It performs best in the regime where the $\chi/3k_{\mathrm{B}} T$ correction is smaller than the $4\lambda k_{\mathrm{B}} T$ factor giving rise to the criterion: $\chi/\lambda < 12 (k_{\mathrm{B}} T)^2$. In the case of the super-ohmic SD used here this becomes simply $ k_{\mathrm{B}} T > \Omega_c $. At lower temperatures, the correction at hand leads to an underestimation of the current in the resonant regime (see Fig.~\ref{Temp_one}) and its overestimation off resonance.
This can be explained as follows:
the energy-dependent hopping rates in the GQME (or the Born-Markov) approach are not symmetric around $\epsilon =  \bar{\varepsilon}_0 \pm \lambda$. Meanwhile, the $\chi/3k_{\mathrm{B}} T$ correction introduced in the Modified Marcus theory induces a symmetric broadening of the Gaussian rates (predicted by MT). For very low temperature, this additional broadening gives rise to the effects discussed above.   

To further investigate the Modified Marcus approach, let us plot the quantity $K(\epsilon) = K^+(\epsilon) + K^-(\epsilon)$, where $K^\pm (\epsilon)$ are given by
\begin{equation}
    K^\pm (\epsilon) = \mathrm{Re}\int_0^\infty \mathrm{d}\tau \: e^{\pm\mathrm{i} \epsilon \tau - \Gamma\tau} B(\tau),
\end{equation}
and are directly proportional to the energy-dependent hopping rates present in Eqs.~(\ref{R1},\ref{R2}). As discussed in Section \ref{comparison}, on resonance, $K(\epsilon)$ directly determines the transport characteristics.
Fig.~\ref{Temp_one}(b) shows $K(\epsilon)$ obtained using the phononic correlation functions as present in the GQME and the Marcus approaches at different temperatures (let us note that the lifetime broadening $\Gamma$ is set to zero in both of the Marcus approaches).
Several aspects of transport characteristics plotted in Fig.~\ref{Temp_one}(a) become more clear.
Firstly, we notice the presence of a peak at $\epsilon = 0$ in the GQME result. It corresponds to an elastic (ground state-to-ground state) electron hopping, and is naturally missing in both of the Marcus approaches which, as the result, significantly underestimate the zero-bias conductance.
Secondly, within the Marcus approach $K^\pm (\epsilon)$ are Gaussian functions centred at $\epsilon = \pm \lambda$. At a given temperature, their width is determined solely by the reorganisation energy. This may give rise to artefacts visible in Fig.~\ref{Temp_one}(a).
On the other hand, within the Modified Marcus approach, the width of these Gaussian functions is also influenced by the parameter $\chi$, which therefore, at intermediate temperatures, removes the spurious effects present in the usual Marcus treatment.

Can the (Modified) Marcus theory be applied to the curcuminoid-based junction from Section \ref{DFT}?
Firstly, the Marcus approaches cannot capture the effects of the structure of the spectral density on the transport characteristics. On a qualitative level, they can therefore only be valid at temperatures at which the structure of the transport characteristics is washed away by the thermal broadening of the Fermi distributions in the leads.
Secondly, the collection of the molecular vibrational modes in Section \ref{DFT} is predominantly of very high frequency -- $\omega_q$ is much larger than $k_{\mathrm{B}}T$ at 300 K for most of the modes. Consequently, the high-temperature assumption of MT will only be satisfied well above  room temperature. Similarly, the criterion for the applicability of the Modified MT yields the temperature of over 600 K.
As the result, the $IV$ characteristics obtained using the GQME approach at room temperature will differ quite markedly from those obtained using the (Modified) Marcus theory.
However, one may expect the Marcus approaches to accurately describe the low-bias transport which is influenced predominantly by the low-frequency vibrational modes (for which the high-temperature assumption of MT is justified). Additionally, the molecular system in the junction can also be coupled to environmental low-frequency modes (of the substrate or the solution in which the junction is immersed). If this coupling dominates, one should expect the room-temperature validity of (M)MT to be recovered.

\subsection{Lifetime-broadened Marcus theory \label{LBMT}}
Let us now return to Eqs. (\ref{R1}, \ref{R2}) and assume that the molecular level interacts predominantly with a relatively low-frequency vibrational environment while $\Gamma$ is comparable to $k_{\mathrm{B}} T$.
We can then take the same limits as in Section \ref{MarcusT} (but retaining the lifetime broadening) to yield:
\begin{widetext}
\begin{align}\label{MGrate1}
& \gamma_l^{\mathrm{LBMT}} =  2\ \Gamma_l \int_{-\infty}^\infty \dfrac{\mathrm{d}\epsilon}{2\pi} f_l(\epsilon)\ \mathrm{Re} \bigg[\sqrt{\dfrac{\pi}{4\lambda k_{\mathrm{B}} T}} \exp\left( \dfrac{(\Gamma - \mathrm{i}\nu_+)^2}{4\lambda k_{\mathrm{B}} T}\right) \mathrm{erfc}\left(\dfrac{\Gamma - \mathrm{i}\nu_+}{\sqrt{4\lambda k_{\mathrm{B}} T}}\right)\bigg] ~;\\     
& \bar{\gamma}_l^{\mathrm{LBMT}} = 2\ \Gamma_l \int_{-\infty}^\infty \dfrac{\mathrm{d}\epsilon}{2\pi} [1 - f_l(\epsilon)]\mathrm{Re} \bigg[\sqrt{\dfrac{\pi}{4\lambda k_{\mathrm{B}} T}} \exp\left( \dfrac{(\Gamma - \mathrm{i}\nu_-)^2}{4\lambda k_{\mathrm{B}} T}\right) \mathrm{erfc}\left(\dfrac{\Gamma - \mathrm{i}\nu_-}{\sqrt{4\lambda k_{\mathrm{B}} T}}\right)\bigg] ~, \label{MGrate2}
\end{align}
\end{widetext}
where, for brevity, $ \nu_{\pm}= \lambda \mp (\epsilon - \bar{\varepsilon}_0)$, and $\mathrm{erfc}(x)$ denotes the complementary error function.
Eqs.~(\ref{MGrate1}, \ref{MGrate2}) form the basis of what we refer to as the \textit{Lifetime-broadened Marcus theory} (LBMT).

Naturally, ignoring the lifetime broadening, by setting $\Gamma = 0$ in Eqs. (\ref{MGrate1}, \ref{MGrate2}), again gives Marcus theory.
On the other hand, by taking the limit $\lambda \rightarrow 0$, one can recover the Landauer--B\"{u}ttiker expression for transport through a single non-interacting level (in a way analogous to what has been done in Appendix \ref{AppA}).
\begin{figure*}
    \centering
    \includegraphics{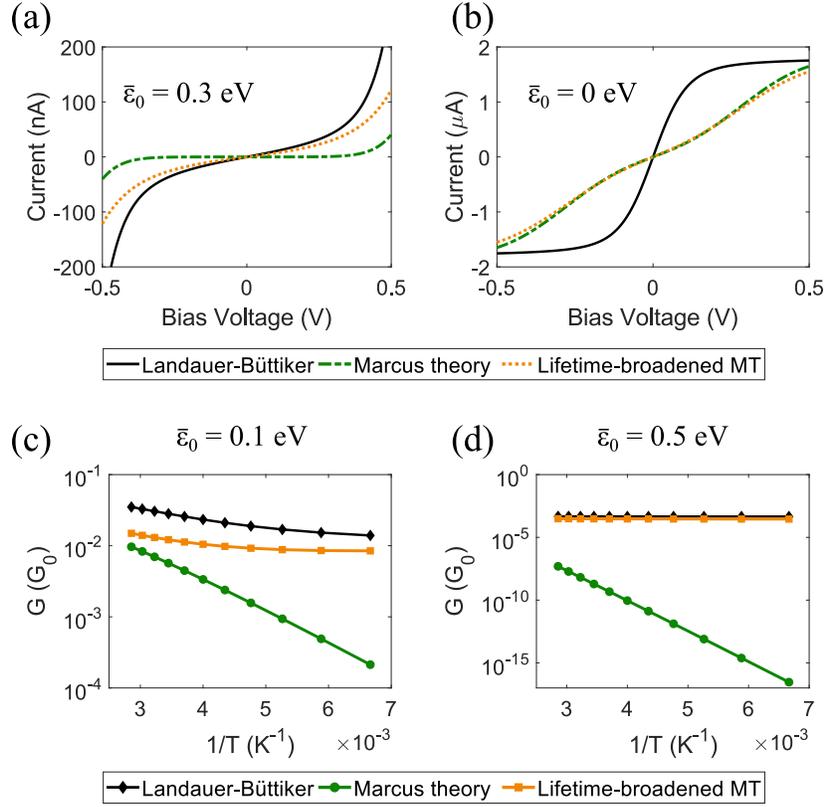}
    \caption{(a,b) $IV$ characteristics obtained using the Landauer--B\"{u}ttiker formalism, Marcus theory and Lifetime-broadened MT for $T = 240$ K, and symmetric molecule-lead coupling: $\Gamma_\mathrm{L} = \Gamma_\mathrm{R} = 15$ meV. (c,d) Zero-bias conductance in units of the conductance quantum, $G_0$. The reorganisation energy in the MT and Lifetime-broadened MT was set to $\lambda = 0.14$ eV. The position of the energy level: (a) $\bar{\varepsilon}_0 = 0.3$ eV; (b) $\bar{\varepsilon}_0 = 0$ eV; (c) $\bar{\varepsilon}_0 = 0.1$ eV; (d) $\bar{\varepsilon}_0 = 0.5$ eV. For consistency the same (renormalised) position of the energy level was used within the Landauer--B\"{u}ttiker approach.}
    \label{figLandauerMarcus}
\end{figure*}

What happens for comparable $\lambda$ and $\Gamma$?
Off resonance ($eV_\mathrm{b}/2 < \bar{\varepsilon}_0$), lifetime-broadened Marcus transport interpolates between the results obtained using Marcus theory and the LB formalism, as shown in Fig. \ref{figLandauerMarcus}(a). In other words, the otherwise elastic Landauer-B\"{u}ttiker transport is suppressed by the vibrational coupling, or conversely, the incoherent Marcus transport is enhanced in the presence of lifetime broadening.   
In the off-resonant regime, transport occurring solely through the hopping mechanism (as described by Marcus theory) yields lower values of current than the purely elastic transport (for given $\Gamma_\mathrm{L}$ and $\Gamma_\mathrm{R}$).
We believe this to be generally the case for the single-site model studied here.
With the molecular energy level within the bias window ($eV_\mathrm{b}/2 > \bar{\varepsilon}_0$), Lifetime-broadened Marcus theory yields lower values of current than the remaining transport mechanisms since it includes all sources of broadening of the $IV$ characteristics (lifetime, phonon, and thermal broadening of the Fermi distributions), see Fig. \ref{figLandauerMarcus}(b).

The temperature dependence of the zero-bias conductance is often used experimentally to identify the transport mechanism.
We consider two cases with the energy level lying $\bar{\varepsilon}_0 = 0.1$ eV and $0.5$ eV above the Fermi energy of the unbiased leads, in the temperature range between 150 and 350 K.
In both cases, the elastic LB transport exhibits a weak temperature dependence (although much more pronounced closer to resonance) stemming from the thermal broadening of the Fermi distributions in the leads.
The temperature dependence of Marcus hopping is dominated by the Arrhenius character of the MT rates giving rise to almost linear dependence of $\mathrm{log}(G/G_0)$ on inverse temperature ($G_0$ is the conductance quantum), see Figs.~\ref{figLandauerMarcus}(c,d).
Perhaps surprisingly, lifetime-broadened Marcus transport does not manifest such  linear behaviour.
Since in the off-resonant regime the elastic (LB) transport through a single-site junction is always more efficient than the incoherent Marcus hopping, even modest amount of lifetime broadening (as incorporated in LBMT) will have a profound effect on the off-resonant conductance.
\begin{figure}
    \centering
    \includegraphics{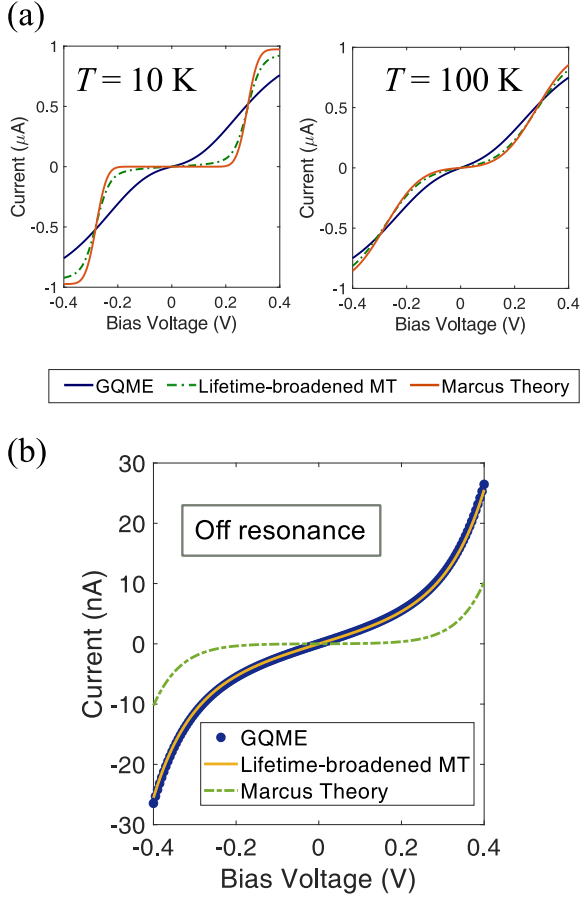}
    \caption{(a) $IV$ characteristics on resonance ($\bar{\varepsilon}_0 = 0$ eV) at different temperatures. 
    (b) Transport characteristics in the off-resonant regime at $T = 300$ K, $\bar{\varepsilon}_0 = 0.3$ eV.
    In both (a) and (b), the molecular energy level is coupled to a vibrational environment with SD of Eq.~\eqref{SD} with $\lambda = 140$ meV and $\Omega_c = 12$ meV; $\Gamma_\mathrm{L} = \Gamma_\mathrm{R} = 8$ meV.}
    \label{Temp_two}
\end{figure}

Let us now repeat the calculation from Fig.~\ref{Temp_one}(a) for the case of lower cut-off frequency $\Omega_{c}$. 
As can be seen in Fig.~\ref{Temp_two}(a), MT gives rise to the same artefacts as in Section \ref{IMarcusT}. Incorporating the lifetime broadening (as is done in LBMT) yields a better agreement with the GQME result, especially at low temperatures.

We also consider the case of off-resonant transport regime at high temperature, see Fig.~\ref{Temp_two}(b). Therein, Lifetime-broadened MT is in an excellent agreement with the GQME result. On the other hand, Marcus theory greatly underestimates the molecular conductance. As shown here, lifetime broadening appears to be a vital part of the transport description off resonance, even at temperatures significantly exceeding $\Gamma$.   

Typically,  resonant transport (as described by the Landauer--B\"{u}ttiker formalism) and  semi-classical electron hopping (Marcus theory) are regarded as two distinct transport mechanisms, see for instance discussion in Ref.~\onlinecite{jia2016covalently}.
As shown here, it is possible to capture both of these phenomena in a simple unifying expression.
It is also apparent that the elastic (Landauer--B\"{u}ttiker) transport and Marcus hopping cannot be considered independently (the total conductance is clearly not a sum of these two contributions).
While  Marcus theory may reasonably well describe the resonant transport regime, its use in the off-resonant case is highly questionable. 

Finally, analogously to what has been done in Section \ref{IMarcusT}, in deriving Eqs.~(\ref{MGrate1}, \ref{MGrate2}) keeping the higher-order term in the expansion of $\coth{\left( {\beta \omega/2} \right)}$  results in expressions equivalent to Eqs.~(\ref{MGrate1}, \ref{MGrate2}) but with $4\lambda k_{\mathrm{B}} T$ terms replaced by $4\lambda k_{\mathrm{B}} T + \chi/3k_{\mathrm{B}} T$. This is what we refer to in Fig.~\ref{scheme1} as the \textit{Modified Lifetime-broadened MT}.

\section{Concluding Remarks \label{end}}
In this work, we derived an expression for current through a molecular junction, modelled as a single electronic level coupled to a bath of thermalised vibrational modes. In appropriate limits it recovers the well-established Marcus theory (Section \ref{MarcusT}),  the perturbative Born-Markov result (Section \ref{BMapp}), as well as the Landauer--B\"{u}ttiker expression for transport through a non-interacting electronic level (Appendix \ref{AppA}).
We have also derived certain extensions of Marcus theory containing: a low-temperature correction (Section \ref{IMarcusT}); lifetime broadening (Section \ref{LBMT}); and both of the above simultaneously (also in Section \ref{LBMT}).
We have shown that Marcus theory description of transport through molecular junctions gives rise to several artefacts: In the resonant tunnelling regime, it often predicts a spurious Franck--Condon blockade and early plateaus in the $IV$ characteristics.
Off resonance, due to the absence of lifetime broadening, it greatly underestimates the molecular conductance.
These issues can be rectified by our here-derived Modified Marcus theory and Lifetime-broadened Marcus theory, respectively.

We have applied our framework to electron-vibrational coupling that was described by both a structureless and a structured spectral density. The former can account for coupling to a broader environment: the solvent\cite{kay2012single,zhang2008single,capozzi2014tunable,choi2016solvent,fatemi2011environmental,milan2015solvent} or the substrate on which the molecule is deposited.\cite{riss2014imaging, fatayer2018reorganization} The latter describes coupling to the intra-molecular vibrational modes and was obtained with the help of \textit{ab initio} calculations. 

We believe that the theoretical framework established here will prove especially  useful in interpreting experimental single-molecule and self-assembled monolayer transport measurements.
Its simplicity should further allow for fitting of the empirical data, and thus extracting various parameters such as the vibrational reorganisation energy and the molecule-lead coupling strengths.

\begin{acknowledgments}
The authors thank James Thomas and Bart Limburg for useful discussions, and N\'{u}ria Aliaga-Alcalde and coworkers for providing us with the results of the DFT calculations from Ref.~\onlinecite{burzuri2016sequential}.
J.K.S. thanks the Clarendon Fund, Hertford College and EPSRC for financial support. E.M.G. acknowledges funding from the Royal Society of Edinburgh and the Scottish Government, J.A.M. acknowledges funding from the Royal Academy of Engineering.
This project was supported by a grant from the John Templeton Foundation. The opinions expressed in this publication are those of the authors and do not necessarily reflect the views of the John Templeton Foundation.
\end{acknowledgments}

\appendix 
\section{Limit of no vibrational coupling \label{AppA}}
It is instructive to consider the expression  \eqref{current} in the limit of zero electron-vibrational coupling: $\mathcal{J}(\omega) = 0$, and consequently $B(t) = 1$. Then,
\begin{equation}
    \gamma_l = 2\ \mathrm{Re}\ \Gamma_l \int_{-\infty}^\infty \dfrac{\mathrm{d}\epsilon}{2\pi} f_l(\epsilon) \int_0^\infty \mathrm{d}\tau e^{\mathrm{i}(\epsilon-\varepsilon_0)\tau} e^{-\Gamma \tau}~,
\end{equation}
and equivalently for $\bar{\gamma}_l$. The one-sided Fourier transform $\Psi(\epsilon) = \mathrm{Re} \int_0^\infty \mathrm{d}\tau \exp{[\left(+\mathrm{i}(\epsilon-\varepsilon_0) -\Gamma\right) \tau]}$ can be easily evaluated:
\begin{equation}
\Psi(\epsilon) = \dfrac{\Gamma}{\Gamma^2 + (\epsilon - \varepsilon_0)^2}~.    
\end{equation}
The numerator in Eq.~\eqref{current} is given by:
\begin{equation}
\gamma_\mathrm{L} \bar{\gamma}_R - \gamma_\mathrm{R} \bar{\gamma}_L = 2 \Gamma_\mathrm{L} \Gamma_\mathrm{R}  \int_{-\infty}^\infty \dfrac{\mathrm{d}\epsilon}{2\pi} [f_L(\epsilon) - f_R(\epsilon)] \Psi(\epsilon) ~,   
\end{equation}
and the denominator simplifies to:
\begin{equation}
\sum_{l = L,R} \gamma_l + \bar{\gamma}_l = \sum_{l = L,R} 2\ \Gamma_l \int_{-\infty}^\infty \dfrac{\mathrm{d}\epsilon}{2\pi} \Psi(\epsilon) = \sum_{l = L,R} \Gamma_l ~.     
\end{equation}
This yields the following expression for the current:
\begin{equation} \label{landauer}
    I = e \int_{-\infty}^\infty \dfrac{\mathrm{d}\epsilon}{2\pi} [f_L(\epsilon) - f_R(\epsilon)] \dfrac{\Gamma_\mathrm{L}\Gamma_\mathrm{R}}{(\epsilon - \varepsilon_0)^2  + \Gamma^2} ~,
\end{equation}
which is the usual Landauer-B\"{u}ttiker expression for a single non-interacting level.\cite{zimbovskaya2013transport} 

\section{NEGF Result \label{AppB}}
The problem considered here has been approached using the non-equilibrium Green function (NEGF) approach by a number of authors.\cite{glazman1988inelastic,wingreen1989inelastic,jauho1994time,galperin2008inelastic,mitra2004phonon,flensberg2003tunneling}
Here, for convenience,  we give the result in the form derived by Jauho \textit{et al.}\cite{jauho1994time}. Using the same notation as in the main body of this work, the electric current is given by:
\begin{equation}\label{NEGF1}
    I = \dfrac{e}{\hbar} \dfrac{\Gamma_\mathrm{L} \Gamma_\mathrm{R}}{\Gamma_\mathrm{L} + \Gamma_\mathrm{R}} \int_{-\infty}^\infty \dfrac{\mathrm{d}\epsilon}{2\pi} [f_L(\epsilon) - f_R(\epsilon)] \int_{-\infty}^\infty \mathrm{d} \tau e^{\mathrm{i}\epsilon \tau}a(\tau) ~,
\end{equation}
where $a(\tau) = \mathrm{i}[G^r(\tau) - G^a(\tau)]$. The retarded Green function is given by:
\begin{equation}
    G^r(\tau) = -\mathrm{i} \theta(\tau) \: e^{-\mathrm{i}\tau \bar{\varepsilon}_0}  e^{- \Gamma \tau} B(\tau) ~.
\end{equation}
In the above, $\theta(\tau)$ is the Heaviside step function, and $G^a (\tau) = [G^r(-\tau)]^\dagger$.

\end{document}